\documentclass[aps,prd,preprint,onecolumn,amsfonts,nofootinbib]{revtex4-2}

\usepackage{amsfonts,amssymb,amsmath,amsthm}
\usepackage{graphicx}
\usepackage{slashed}
\usepackage{hyperref}
\newcommand{\Ltwi}{\widetilde{L}}
\newcommand{\dd}{\mathrm{d}}
\newcommand{\ii}{\mathrm{i}}
\newcommand{\MM}{\mathcal{M}}

\begin{document}
\begin{titlepage}
\title{Carroll limit of one-loop effective action}
\author{Dmitri Vassilevich}\email{dvassil@gmail.com}
\affiliation{Centro de Matem{\'a}tica, Computa\c{c}\~{a}o e Cogni\c{c}\~{a}o - Universidade Federal do ABC, Avenida do Estados 5001, CEP 09210-580, Santo Andr\'{e}, SP, Brazil}

\begin{abstract}
In this paper, we consider a Carroll magnetic limit of a one-loop scalar effective action. We work on general static backgrounds and compute both divergent and finite parts of the effective action in this limit. We show, that the divergent part can be removed by adding local counterterms. The finite part is related to an effective action in a lower dimensional theory which however does not coincide in general with the one obtained by a Carroll limit in the classical counterpart. 
\end{abstract}

\maketitle

\end{titlepage}

\section{Introduction}\label{sec:intro}
Carroll symmetries were introduced long ago \cite{LevyLeblond1965,SenGupta:1966qer} but they received due attention much later after the discovery of corresponding conformal symmetries \cite{Duval:2014uoa,Duval:2014uva,Duval:2014lpa}. Over the past few years considerable progress in understanding of various aspects of Carroll theories has been achieved. To give just a few examples, we mention the works \cite{Ciambelli:2018wre,Figueroa-OFarrill:2021sxz,Herfray:2021qmp} on the Carroll structure at null infinity, and on propagating Carroll fields (Carroll swiftons) \cite{Ecker:2024czx}.

Several works were dedicated to quantum Carroll theories \cite{Bagchi:2016bcd,Donnay:2022aba,Bagchi:2022emh,Donnay:2022wvx,Chen:2021xkw,Mehra:2023rmm,Banerjee:2023jpi,Chen:2024voz,Cotler:2024xhb,Ruzziconi:2024zkr} mostly studying quantization, Carrollian conformal field theories, and  correlation functions, but also quantum Hall effect \cite{Marsot:2022imf}, representation structure and thermodynamics \cite{Figueroa-OFarrill:2023qty}, renormalisation group flow as an origin of Carrollian symmetries \cite{Bagchi:2024unl}, and Hawking radiation \cite{Aggarwal:2024yxy}. It was shown, at some simple examples, that in Carroll limit the scalar partition function  is divergent \cite{deBoer:2023fnj}.

The purpose of present work is to perform a detailed analysis of divergent and finite terms in the Carroll (magnetic) magnetic limit of the one-loop effective action in scalar theories in various dimensions. We will consider static but otherwise quite general backgrounds.

Carroll limit is basically an ultrarelativistic limit $c\to 0$, where $c$ is the speed of light. 
Carroll geometry is defined through a vector $v^\mu$ which, roughly speaking, fixes the time direction, and a spatial metric $h_{\mu\nu}$. Together with the fields $\tau_\mu$ and $h^{\mu\nu}$, they satisfy the relations
\begin{equation}
v^\mu h_{\mu\nu}=0,\quad \tau_\mu h^{\mu\nu}=0,\quad v^\mu \tau_\mu =-1,\quad
h_\mu^\nu \equiv h_{\mu\rho}h^{\rho\nu}=\delta_\mu^\nu + v^\nu \tau_\mu.\label{hvtau}
\end{equation}  
Local Carroll boosts with the parameter $\Lambda_\mu$ act as follows
\begin{equation}
\delta_\Lambda v^\mu=0,\quad \delta_\Lambda \tau_\mu = \Lambda_\mu,\quad
\delta_\Lambda h^{\mu\nu}=\bigl( h^{\mu\rho}v^\nu + h^{\nu\rho}v^\mu\bigr) \Lambda_\rho,\quad \delta_\Lambda h_{\mu\nu}=0.\label{Cboosts} 
\end{equation}
The parameter $\Lambda$ satisfies the restriction $\Lambda_\mu v^\mu=0$. It is easy to check that the volume element 
\begin{equation}
e=\bigl( \det (\tau_\mu\tau_\nu +h_{\mu\nu})\bigr)^{1/2} \label{vol}
\end{equation}
is invariant under Carroll boosts.

In this work, we will be interested in scalar field theories. The so called
electric and magnetic Carroll scalar theories were defined as contractions from Lorentz-invariant theories in \cite{Henneaux:2021yzg} (see also an earlier work \cite{Bergshoeff:2014jla}) while
conformal Carroll scalar theories were constructed in \cite{Rivera-Betancour:2022lkc,Baiguera:2022lsw}. We will consider magnetic theories only. The action for magnetic scalar on an $n+1$ dimensional manifold $\MM$ reads
\begin{eqnarray}
I[\Pi,\phi]=\int_{\MM} \dd x^{n+1} e\, \left( \Pi v^\mu \partial_\mu \phi + h^{\mu\nu}\partial_\mu\phi\, \partial_\nu \phi +V(\phi) \right).\label{IPp}
\end{eqnarray}
We work in the Euclidean signature. This action is invariant under local Carroll boosts if the rules (\ref{Cboosts}) are supplemented by a suitable transformation rule for $\Pi$. $V$ is a potential. The field $\Pi$ generates a
constraint
\begin{equation}
v^\mu \partial_\mu\phi =0. \label{constr}
\end{equation}
With this constraint, the action becomes
\begin{eqnarray}
I[\phi]=\int_{\MM} \dd x^{n+1} e\, \left(  h^{\mu\nu}\partial_\mu\phi\, \partial_\nu \phi +V(\phi) \right). \label{Iphi}
\end{eqnarray}
This action is Carroll boost invariant provide $\phi$ satisfies (\ref{constr}).

Consider an ``electromagnetic" scalar action \cite{Ciambelli:2023xqk}
\begin{eqnarray}
I_{\mathrm{em}}[\phi]=\int_{\MM} \dd x^{n+1} e\, \left( \kappa (v^\mu \partial_\mu \phi)^2 + h^{\mu\nu}\partial_\mu\phi\, \partial_\nu \phi + V(\phi) \right),\label{Iem}
\end{eqnarray}
where $\kappa$ is a positive coupling constant having the meaning of $1/c^2$. In the limit $\kappa\to\infty$ the time variation of $\phi$ are suppressed so that one gets the dynamics described by (\ref{Iem}) with the constraint(\ref{constr}). The purpose of present work is to see what happens in quantum theory in this limit.

Since the $\kappa\to\infty$ limit means imposing the constraint (\ref{constr}), one may expect that in this limit also the quantum effective action will be defined through an $n$ dimensional theory living on a hypersurface of constant $y$, where $y$ is defined through $\partial_y=v^\mu\partial_\mu$. However, in this singular limit the determinant of $n+1$ dimensional operator may keep the memory on the way the constant $y$ hypersurfaces are embedded in $\MM$ and on the function $v^\mu$. (This happens, e.g., with Faddeev--Popov determinants in noncovariant gauges, see \cite{Vassilevich:1994cz}.) Moreover, even though the contribution of each $y$-dependent mode is suppresses at the $\kappa\to\infty$ limit, there are infinitely many such modes. Their collective contribution may be non-vanishing and even divergent.

For a generic scalar field theory the one-loop effective action is given by 
\begin{equation}
W(L)=\tfrac 12 \ln \det L \label{lndetL}
\end{equation}
where $L$ is an operator of Laplace type appearing in the quadratic form of classical action.
We use the $\zeta$ function regularization and write the regularized determinant as
\begin{equation}
(\ln \det L)_s=-\Gamma(s) \zeta(s,L), \label{zetareg}
\end{equation}
where $s$ is a complex regularization parameter and 
\begin{equation}
\zeta(s,L)=\mathrm{Tr}\, (L^{-s}) \label{defzeta}
\end{equation}
is the spectral $\zeta$ function of $L$. The regularized effective is $W_s=\tfrac 12 (\ln \det L)_s$. The physical limit corresponds to an analytic continuation to the point $s=0$. Near this point,
\begin{equation}
\Gamma(s)\zeta(s,L)\simeq \left( \tfrac{1}{s} - \gamma_{\mathrm{E}}\right)\zeta(0,L)+\zeta'(0,L)+\mathcal{O}(s) \label{nears0} 
\end{equation}
where $\gamma_{\mathrm{E}}$ is the Euler constant.

The heat kernel expansion \cite{Vassilevich:2003xt} will be a useful tool.
If $f$ is a smooth function, there is an asymptotic expansion of the smeared heat kernel at $t\to +0$,
\begin{equation}
K(t,L)=\mathrm{Tr}\, \left( f\, e^{-tL}\right)\simeq \sum_{k=0}^\infty t^{\frac{k-m}{2}} a_k(f,L)\label{heatk}
\end{equation}
Here $m$ is dimension of the base manifold. For $f=1$ we will use a shorthand notation $a_k(L)\equiv a_k(1,L)$. Because of the relation 
\begin{equation}
\zeta(0,L)=a_m(L)\label{zetam}
\end{equation}
the pole term in (\ref{nears0}) is given by $a_m(L)$. 

This work is organized as follows. We consider a scalar theory on a product manifold $\MM=\widetilde{\MM}\times S^1$ with static background fields. We define an operator $\bar{L}$ acting on quantum fluctuations and apply a local scale transformation to transform this operator to another operator $\Ltwi$. This latter operator is a sum of two commuting operators, one being independent of $\kappa$ and the other having an exactly known spectrum. The difference between effective actions for $\bar{L}$ and $\Ltwi$ is given by scale anomaly. These steps are done in Section \ref{sec:scale}. The $\kappa$ expansion of the effective action for $\Ltwi$ is constructed in Section \ref{sec:ex} where the divergences in this expansion are analysed. Our most important observation is that all divergences in the $\kappa\to \infty$ expansion may be removed by local counterterms. A simple example in two dimensions is considered in Section \ref{sec:ex}. The last Section \ref{sec:dis} contains concluding remarks.

\section{The Carroll limit}
\subsection{Local scale transformation}\label{sec:scale}
We take $\MM$ being a product manifold, $\MM=\widetilde{\MM}\times S^1$,  and coordinates $x^\mu=(z^j,y)$ such that $y$ is a coordinate on unit $S^1$ and $z^j$ parametrize $\widetilde{\MM}$. The ``temporal" direction is taken along $S^1$, $v^\mu=v^y$. The components $h_{\mu\nu}$ with $\mu$ or $\nu$ in the $S^1$ direction vanish while the components $h_{ij}$ are identified with a Riemannian metric on $\widetilde{\MM}$. $h^{ij}$ is taken to be an inverse of $h_{ij}$, $h_{ij}h^{jk}=\delta_i^k$. This choice breaks Carroll boost symmetry. We will discuss the consequences below. The metric $h^{ij}$ and $v^y$ are assumed to be static, i.e. these fields depend on $z$ only. For technical reasons we also assume that $v^y$ is non-vanishing while $h^{ij}$ is non-degenerate. This excludes configurations with (Euclidean) horizons, like Carrollian Rindler spacetimes \cite{Li:2024kbo}, for example.

We represent $\phi=\Phi+\varphi$ with $\Phi$ being a static background field and $\varphi$ being a quantum fluctuation. By expanding (\ref{Iem}) in powers of $\varphi$ and keeping quadratic terms only, we obtain
\begin{equation}
I_{\mathrm{em}}^{(2)}=\int_{\MM} \dd x^{n+1} e\, \varphi \, \bar{L}\varphi \label{Iem2}
\end{equation} 
where 
\begin{equation}
\bar{L}=-\bar{g}^{\mu\nu}\bar{\nabla}_\mu\bar{\nabla}_\nu +\xi \bar{R} +\bar{U}\label{barL}
\end{equation}
Here $\bar{g}^{\mu\nu}=(h^{ij},\kappa (v^y)^2)$ is an effective metric. $\bar\nabla$ and $\bar{R}$ are the covariant derivative and the Riemannian curvature for this metric, respectively. One can check that $\bar{R}$ does not depend on $\kappa$. For convenience, we separated a term with conformal coupling in $n+1$ dimensions, $\xi=\tfrac{n-1}{4n}$, so that $\bar{U}=-\xi\bar{R}+\tfrac 12 V''(\Phi)$.

A diffeomorphism and Carroll boost invariant path integral measure reads\footnote{Form the operator theory point of view, it is more natural to have in both (\ref{Iem2}) and (\ref{measure}) the volume element $\sqrt{\bar{g}}=\kappa^{-1/2}e$ instead of $e$ since $\bar{g}$ is the metric appearing in the leading symbol of $\bar{L}$. However, removing the multiplier $\kappa^{-1/2}$ at both places does not change the effective action. }
\begin{equation}
\int \mathcal{D}\varphi\, \exp \left( -\int \dd^{n+1}x\, e\, \varphi^2\right)=1
\label{measure}
\end{equation}
Taking the path integral in the Gaussian approximation one arrives at a one-loop effective action $W(\bar{L})$, see (\ref{lndetL}).

We have fixed $h^{\mu\nu}$ to be a metric over $\widetilde{\MM}$ to facilitate taking the $\kappa\to\infty$ limit, see below. Let us lift this restriction for a moment, perform an infinitesimal Carroll boost (\ref{Cboosts}) and check invariance of the effective action $W(\bar{L})$. Let us suppose that the couplings contained in $V(\phi)$ are invariant under these transformations. Such couplings may be $\phi^k$ and $\bar{R}\phi^2$, for example. The volume element $e$ and hence the path integral measure are invariant. The variation of $\bar{L}$ reads
\begin{equation}
\delta_\Lambda \bar{L}=-2\Lambda_\mu h^{\mu\nu} v^{\sigma} \bar{\nabla}_\nu\bar{\nabla}_\sigma.\label{delLamL}
\end{equation}
The change of sign $\Lambda_\mu\to -\Lambda_\mu$ in (\ref{delLamL}) can be be compensated by a reflection $v^\mu\to -v^\mu$ leaving $\bar{L}$ unchanged. This means that
\begin{equation}
\delta_\Lambda W(\bar{L})=\tfrac 12 \mathrm{Tr}\, \bigl( \delta_\Lambda \bar{L} \cdot \bar{L}^{-1}\bigr) \label{infboostW}
\end{equation}
is invariant under $\Lambda_\mu \to -\Lambda_\mu$. Thus, the infinitesimal Carroll boost of effective action (\ref{infboostW}) vanishes.

Let us denote denote $v^y\equiv \exp(\rho)$ and make a local scale transformation of the operator $\bar{L}$. 
\begin{equation}
\bar{L}=e^{\frac{n+3}{2} \rho}\, \Ltwi \, e^{-\frac{n-1}{2}\rho},\label{trafoL}
\end{equation}
where
\begin{equation}
\Ltwi =-\kappa \partial^2_y +L_n \label{Ltdef}
\end{equation}
and
\begin{eqnarray}
&&L_n=-\nabla_j\nabla^j +\xi R +e^{-2\rho}\bar{U} \nonumber\\
&&\quad = -\nabla_j\nabla^j -\tfrac 14 (n-1)\left(2\nabla^2\rho +(1-n) (\nabla\rho)^2 \right)+\tfrac 12 e^{-2\rho} V''(\Phi).\label{Ln}
\end{eqnarray}

We stress, that the transformation (\ref{trafoL}) does not coincide with local conformal (Weyl) transformations of the metric since $\bar{L}$ is not necessarily Weyl covariant. Thus, usual Weyl transformation of the metric $\bar{g}^{\mu\nu}= e^{2\rho}g^{\mu\nu}$ is accompanied by a local rescaling of the potential. In Eq.\ (\ref{Ln}) all covariant derivatives $\nabla_j$ are the Riemannian derivatives with the metric $h_{ij}$ on $\widetilde{\MM}$. $R$ is the curvature of $h_{ij}$.

Let us consider a family of operators
\begin{equation}
L_u=e^{\frac{n+2}{2}u \rho}\, \Ltwi \, e^{-\frac{n-2}{2}u\rho}
\end{equation}
with $u\in [0,1]$, so that $L_1=\bar{L}$ and $L_0=\Ltwi$. The derivative of $\zeta$ function with respect to $u$ reads
\begin{equation}
\frac{\dd}{\dd u}\, \zeta(s,L_u)=-2s \mathrm{Tr} \left( \rho L_u^{-s}\right) \label{dduz}
\end{equation}
Due to a factor of $s$ in (\ref{dduz}) the variation of regularized effective action is finite at $s=0$. 
By applying an analog of Eq. (\ref{zetam}), see \cite{GilkeyNew,Vassilevich:2003xt}, we obtain
\begin{equation}
\frac{\dd}{\dd u}\, W_s\vert_{s=0}=a_{n+1}(\rho,L_u) \label{ddWs}
\end{equation}
The heat kernel coefficients are integrals of local invariants constructed from the symbol of operator $L_u$, see \cite{Vassilevich:2003xt}. By changing the coordinate $y\to y'=\kappa^{-1/2}y$ one removes the dependence of integrand on $\kappa$. The whole dependence of $a_{n+1}(\rho,L_u)$ on $\kappa$ resides in the size of integration interval of $y'$. Thus, $a_{n+1}(\rho,L_u)\propto \kappa^{-1/2}$ and
\begin{equation}
\lim_{\kappa\to\infty}a_{n+1}(\rho,L_u)=0.
\end{equation}
Thus we conclude that in the Carroll limit $\kappa\to\infty$ the effective action for operator $\bar{L}$ equals to the effective action for $\Ltwi$.

\subsection{Calculation of the $\kappa$-expansion}\label{sec:cal}
The operator $\Ltwi$ is a sum of two commuting operators, $\kappa\partial_y^2$ and $L_n$. The spectrum of the former is known exactly, while the second one does not depend on $\kappa$. This facilitates calculation of  the large $\kappa$ expansion of the effective action. 
We will evaluate this expansion by doing small adjustments and modification in the method proposed in \cite{Dowker:1978md,Dowker:1983ci} to calculate the high temperature expansion of free energy. We write
\begin{eqnarray}
&&\zeta(s,\Ltwi)=\frac{1}{\Gamma(s)}\int_0^\infty \dd t\, t^{s-1} K(t,\Ltwi)=\nonumber\\
&&\qquad =\frac{1}{\Gamma(s)}\int_0^\infty \dd t\, t^{s-1} \sum_{l=-\infty}^\infty
e^{-t\kappa l^2} K(t,L_n) \nonumber\\
&&\qquad = \frac{1}{\Gamma(s)}\int_0^\infty \dd t\, t^{s-1} \vartheta\left(0,\ii t \kappa \pi^{-1}\right) K(t,L_n)\nonumber\\
&&\qquad =\zeta(s,L_n) + \frac{1}{\Gamma(s)}\int_0^\infty \dd t\, t^{s-1}\left[ \vartheta\left(0,\ii t \kappa \pi^{-1}\right)-1\right] K(t,L_n) .\label{zetaLt}
\end{eqnarray}
Here we used the Mellin transform to represent $\zeta$ function through the heat kernel. Then, we separated $\kappa\partial_y^2$ from that of $L_n$ and substituted exact spectrum of the former operator. The
Jacobi $\vartheta$ function
\begin{equation}
\vartheta(z,\tau)=\sum_{l=-\infty}^\infty \exp (\ii\pi l^2\tau +2\ii \pi l z) \label{Jactheta} 
\end{equation}
is used to calculate the sum over $l$.

Let us substitute in (\ref{zetaLt}) the heat kernel expansion (\ref{heatk}) for $K(t,L_n)$. The integration over $t$ is performed with the help of Riemann formula for the Mellin transform 
\begin{equation}
\frac 12 \int_0^\infty \dd t\, t^{\sigma -1} \left[\vartheta(0,\ii t) - 1\right] = \Gamma (\sigma) \pi^{-\sigma} \zeta_{\mathrm{R}}(2\sigma). \label{Riem}
\end{equation}
Here $\zeta_{\mathrm{R}}$ is the Riemann $\zeta$ function.

\begin{equation}
\zeta(s,\Ltwi)=\zeta(s,L_n)+\frac{2}{\Gamma(s)}\sum_{k=0}^\infty a_k(L_n) \, \Gamma \left( \frac{k-n}{2}+s \right)\, \zeta_{\mathrm{R}}\left( k-n +2s\right)\, \kappa^{-s +\frac{n-k}{2}}. \label{expzeta}
\end{equation}
Let us substitute this expansion in the formulas (\ref{zetareg}) and (\ref{lndetL}) to obtain an expansion for regularized effective action $W_s$. Non-negative powers of $\kappa$ in this expansion contain two pole terms, one coming from $\zeta(s,L_n)$ and the other -- from the term with $a_n(L_n)$ in the sum. Due to the identity (\ref{zetam}), these two poles cancel each other, so that we can immediately take the limit $s\to 0$. 
\begin{eqnarray}
&&W=-\sum_{p=1}^{\lfloor n/2 \rfloor} \pi^{-2p} \frac{(2p-1)!!}{2^p}\, \zeta_{\mathrm{R}}(2p+1)\, a_{n-2p}(L_n) \kappa^p \nonumber\\
&&\quad +\sum_{p=1}^{\lfloor (n+1)/2 \rfloor}\frac{(-1)^p B_{2p}2^{p-1}}{(2p-1)!!\, p}\, a_{n-2p+1}(L_n)\, \kappa^{p -\frac 12}\nonumber\\
&&\quad -\frac 12 \zeta'(0,L_n) +\frac 12 \ln (2\pi\kappa)\, a_n(L_n) +\mathcal{O}(\kappa^{-1/2}).\label{kappaexp}
\end{eqnarray}
Here $B_p$ are the Bernoulli numbers. The only surviving pole term in (\ref{expzeta}) is proportional to $\kappa^{-1/2}$ and reads
\begin{equation}
W_{s,\mathrm{pole}}= -\frac{1}{2s} \sqrt{\frac{\pi}{\kappa}}\, a_{n+1} (L_n) \label{pole}
\end{equation}
Due to the product structure of operator $\Ltwi$ there is a simple relation between the heat kernel coefficients 
\begin{equation}
a_k(\Ltwi)=\sqrt{\frac{\pi}{\kappa}}\, a_k(L_n) \label{akak}
\end{equation}
valid for any $k$.
Thus, the pole term (\ref{pole}) is nothing else than the standard ultraviolet divergence in the effective action for $\Ltwi$, see (\ref{nears0}) and (\ref{zetam}).

On manifolds without boundaries all odd-numbered heat kernel coefficients vanish. Thus, for even (respectively, odd) $n$ only the first (respectively, the second) sum survives in the expansion (\ref{kappaexp}). For the same reason, the pole term (\ref{pole}) vanishes identically already for finite values of $\kappa$ if $n+1$ is odd. This reflects the fact that there are no divergences in the $\zeta$ function regularization in odd dimensions. In the the $\kappa\to\infty$ limit, this term vanishes for all $n$. Instead, also for all $n$ the terms which are divergent in the Carroll limit appear. Since all these terms are given by heat kernel coefficients, they are local. Thus, \emph{the one loop effective action can be made finite in the Carroll limit by subtracting local counterterms.} This is the main result of this work. 

There is a relation between divergences in the Carroll limit and ultraviolet divergences in the parent theory (\ref{Iem}) with $\kappa=1$. In the spectral cutoff regularization (which is equivalent to changing the lower limit of integration in the Mellin transform of formulas (\ref{zetaLt}) to $\Lambda^{-2}$ with $\Lambda$ being a cutoff scale) the power-law divergences are $\propto a_k(\bar L(\kappa=1))\Lambda^{n+1-k}$, $k=0,\dots,n$, see Eq.\ (1.21) in \cite{Vassilevich:2003xt}. Due to the reduction formula (\ref{akak}) with $\kappa=1$, the counterterms needed to remove the divergences in the Carroll limit are the same as power-law divergences in the theory described by the scale transformed operator $\Ltwi(\kappa=1)$. If we wish to discuss renormalizability, we should restrict ourselves to non-gravitational backgrounds allowing for constant $v^\mu$ and $h^{\mu\nu}$. Since a constant $v^\mu$ can be absorbed in $\kappa$, the local scale transformation (\ref{barL}) becomes trivial. In this case, the counterterms needed to remove the $\kappa\to \infty$ divergences are the same as the counterterms needed in the in the parent theory to remove power-law divergences. In other words, in a renormalizable theory the counterterms appearring in the Carroll limit do not destroy renormalizability. The same conclusion can be reached by directly counting canonical mass dimensions of the structures appearing in $a_k(L_n)$.

The partition function magnetic scalar was constructed in \cite{Cotler:2024xhb} basing on an interpretation through an $n$ dimensional statistical system. No regularization was applied in \cite{Cotler:2024xhb}. In the $\zeta$ function regularization the finite part of corresponding effective action coincides with finite terms in (\ref{kappaexp}).

\subsection{An example}\label{sec:ex}
To be more explicit, let us consider an example of Carroll limit in a two-dimensional theory, $n=1$. Take $\MM=S^1\times \widetilde{\MM}$ where $\widetilde{\MM}$ is one-dimensional. Let us denote $h^{zz}:=e^{2\psi(z)}$. With $V=m^2\phi^2$ the action (\ref{Iem}) takes the form
\begin{equation}
I_{\mathrm{em}}=\int_{\MM}\dd^2x\, e^{-\rho-\psi}\bigl( \kappa e^{2\rho}(\partial_y\phi)^2 + e^{2\psi} (\partial_z\phi)^2 +m^2\phi^2\bigr).\label{n1Iem}
\end{equation}
After integrating by parts in this expression we arrive at
\begin{equation}
\bar{L}=e^{2\rho}\left( -\kappa \partial_y^2 - e^{2(\psi-\rho)} (\partial_z +
(\partial_z\psi -\partial_z\rho))\partial_z +e^{-2\rho}m^2 \right). \label{n1barL}
\end{equation}
Through a local scale transformation, $\bar{L}=e^{2\rho}\Ltwi$, cf. (\ref{trafoL}) and (\ref{Ln}), this operator defines 
\begin{equation}
L_1=- e^{2(\psi-\rho)} (\partial_z +
(\partial_z\psi -\partial_z\rho))\partial_z +e^{-2\rho}m^2.
\label{n1L1}
\end{equation}
Just few terms remain in the expansion (\ref{kappaexp}),
\begin{equation}
W=-\tfrac 16 a_0(L_1) \kappa^{1/2}-\tfrac 12 \zeta'(0,L_1) +\mathcal{O}(\kappa^{-1/2}).\label{n1Wexp}
\end{equation}
The heat kernel coefficient $a_0(L_1)$ can be easily computed (see \cite{Vassilevich:2003xt}, e.g.),
\begin{equation}
a_0(L_1)=(4\pi)^{-1/2} \int_{\widetilde{\MM}}\dd z\, e^{\rho-\psi} 
\label{a0L1}
\end{equation}
which is local, as expected. The finite term, $\tfrac 12 \zeta'(0,L_1)$ is a one-loop effective action in a one-dimensional theory. This term is nonlocal, though its structure is considerably simpler than that in higher dimensional theories
\cite{Kirsten:2001wz}. If $\widetilde{\MM}$ is a unit $S^1$ and $\rho=\psi=0$, the eigenvalues of $L_1$ are $k^2+m^2$. At a very formal level, one may write $-\tfrac 12 \zeta'(0,L_1)$ as $\tfrac 12 \sum_{k\in \mathbb{Z}} \ln (k^2+m^2)$. The latter expression is, of course, divergent. It is also consistent with an expression for magnetic limit of the scalar partition function in \cite{deBoer:2023fnj} where no regularization was used.

For $m=0$, the operator (\ref{n1L1}) depends on $\rho$ and $\psi$ through the combination $\rho-\psi$ only. Thus, $L_1$ is invariant under local conformal transformations of the metric $\bar{g}^{\mu\nu}$. 

\section{Discussion and conclusions}\label{sec:dis}
In this paper, we considered a Carroll magnetic limit $\kappa\to\infty$ of the one-loop effective action in an $n+1$ dimensional scalar theory. The coefficients in $\kappa$ expansion are expressed through spectral characteristics (heat kernel coefficients and $\zeta'(0)$) of an $n$ dimensional operator $L_n$. The operator $L_n$ is not the kinetic operator $\bar{L}$ of the original theory with the term $(v^\mu\partial_\mu)^2$ neglected, as one might have expected naively, but rather an operator obtained through a local scale transformation thereof. We observed, that the usual $1/s$ divergence of $\zeta$ regularization vanishes in the $\kappa\to\infty$ limit. Instead, several divergent terms with positive powers of $\kappa$ or with $\ln(\kappa)$ appear. All of them are local. Thus, the effective action can be made finite in Carroll limit by adding a finite number of local counterterms.

Our starting point, the action (\ref{Iem}), is not Carroll boost invariant for finite $\kappa$. Besides, we fixed $h^{\mu\nu}$ to be an (inverse) metric over $\widetilde{\MM}$ thus breaking the Carroll boost invariance. Somewhat surprisingly, the effective action appeared to be invariant under infinitesimal Carroll boosts, see discussion around Eq.\ (\ref{infboostW}). It may be, that this invariance is accidental following from a rather specific choice of the background. A more thorough study of Carroll boost invariance requires lifting the assumption made above and considering generic $h^{\mu\nu}$. Such analysis will be considerably more difficult than the one presented in this work.  Technical tools like the ones which have been developed for high temperature limits on stationary (rather than static) backgrounds \cite{Fursaev:2001yu} will be useful.

A word of warning is in order. The $\kappa$ expansion of effective action has to be used with care. Since the effective metric $g$ depends on $\kappa$, the integrated quantities (like the effective action) can be of a different order in $\kappa$ than corresponding local quantities (like the stress energy tensor). Besides, the components of tensors with upper and lower indices can contain different powers of $\kappa$. An example of such situation can be found in the paper \cite{Aggarwal:2024yxy} where the Hawking effect on for a 2D Carroll--Schwarzschild black hole \cite{Ecker:2023uwm} was studied. As usual for massless two-dimensional theories \cite{Christensen:1977jc}, the whole information on Hawking effect is contained in the scale anomaly, which gives a contribution to the effective action suppressed by $\kappa^{-1/2}$. However, the Carroll limit of stress energy tensor contained both finite and divergent contributions. In other words, to analyse local quantities produced by taking variation derivatives of the effective action one may need to use more terms in the $\kappa$ expansion. Such terms are easily obtained by extending the summation ranges in the sums (\ref{kappaexp}). 

As well as vacuum expectation values of the components stress energy tensor, the $N$-point functions may also be divergent in the Carroll limit despite renormalizability of the effective action. In a different context, a similar effect was observed in \cite{Capri:2006bj} where Yang--Mills theories in gauges interpolating between Landau and Coulomb were studied. Although the theory was renormalizable at the level of the action, $N$-point function exhibited a singular behaviour in the Coulomb limit.

An extension of our results to more general and even $\kappa$-dependent interaction $V(\phi)$ is straightforward. It is sufficient to re-expand all terms in (\ref{kappaexp}) in powers of $\kappa$. This is easy to do with the heat kernel coefficients, but may be more tedious in the case of $\zeta'(0,L_n)$. An expansion of the effective action for the fields of other spins can be obtained along similar lines. It seems promising to apply our methods to the $c\to 0$ limit in Conformal Field Theories \cite{Fiorucci:2023lpb} and to quantum theories where only external lines of Feynman diagrams are Carrollian \cite{Liu:2024nfc}.

\begin{acknowledgments}
I am grateful to Ankit Aggarwal, Florian Ecker, and Daniel Grumiller for discussions and collaboration on related topics. I also thank the Erwin-Schr\"{o}dinger Institute (ESI) for the hospitality in April 2024
during the program “Carrollian physics and holography” and I am grateful to the participants of this program for numerous insightful discussions.
This work was supported in parts by the S\~ao Paulo Research Foundation (FAPESP) through the grant 2021/10128-0, and by the National Council for Scientific and Technological Development (CNPq), grant 304758/2022-1. 
\end{acknowledgments}

\bibliography{carroll}

\end{document}